\documentclass[pdftex]{article}
\usepackage{spconf,amsmath,graphicx}
\usepackage{cite,url,booktabs,subcaption}

\title{PeriodNet: A non-autoregressive waveform generation model with a structure separating periodic and aperiodic components}
\name{Yukiya Hono, Shinji Takaki, Kei Hashimoto, Keiichiro Oura, Yoshihiko Nankaku, and Keiichi Tokuda}
\address{Department of Computer Science, Nagoya Institute of Technology, Nagoya, Japan}
\begin{document}
\ninept
\maketitle
\begin{abstract}
We propose PeriodNet, a non-autoregressive (non-AR) waveform generation model with a new model structure for modeling periodic and aperiodic components in speech waveforms.
The non-AR waveform generation models can generate speech waveforms parallelly and can be used as a speech vocoder by conditioning an acoustic feature.
Since a speech waveform contains periodic and aperiodic components, both components should be appropriately modeled to generate a high-quality speech waveform.
However, it is difficult to decompose the components from a natural speech waveform in advance.
To address this issue, we propose a parallel model and a series model structure separating periodic and aperiodic components.
The features of our proposed models are that explicit periodic and aperiodic signals are taken as input, and external periodic/aperiodic decomposition is not needed in training.
Experiments using a singing voice corpus show that our proposed structure improves the naturalness of the generated waveform.
We also show that the speech waveforms with a pitch outside of the training data range can be generated with more naturalness.

\end{abstract}
\section{Introduction}
\label{sec:intro}
\vspace{-1mm}
In recent years, speech synthesis technology has rapidly improved with the introduction of deep neural networks.
In particular, WaveNet~\cite{oord-2016-wavenet}, which has an autoregressive (AR) structure, directly models the distributions of waveform samples and has demonstrated remarkable performance.
WaveNet can be used as a speech vocoder by conditioning auxiliary features such as Mel-spectrogram and acoustic features extracted by conventional signal processing-based vocoder~\cite{tamamori-2017-speaker}.
This is also used in state-of-the-art speech synthesis systems, which greatly contributes to improving the quality of synthesized speech~\cite{shen-2018-natural,ping-2017-deep}.
However, WaveNet suffers from slow inference speed because of the AR mechanism and huge network architectures.
Although compact AR models~\cite{kalchbrenner-2018-efficient,valin-2019-lpcnet} have been proposed to accelerate inference speed, it is limited because audio samples must be generated sequentially.
Thus, such models are not suited for real-time TTS applications.

Recently, significant efforts have been devoted to building non-AR models to resolve this problem.
Parallel WaveNet~\cite{oord-2018-parallel} and ClariNet~\cite{ping-2018-clarinet} introduce the teacher-student knowledge distillation.
This framework transfers the knowledge from an AR teacher WaveNet to an inverse autoregressive flow (IAF)-based non-AR student model~\cite{kingma-2016-improved}.
The IAF student model is highly parallelizable and can synthesize high-quality waveforms.
However, the training procedure is complicated because it requires a well-trained teacher model as well as a mix of distilling and other perceptual training criteria.
WaveGlow~\cite{prenger-2019-waveglow} and FloWaveNet~\cite{kim-2018-flowavenet} with flow-based generative models have been proposed as well.
Although these models can be directly learned by minimizing the negative log-likelihood of the training data, they need a huge number of parameters and require many GPU resources to obtain optimal results for a single speaker model.

Another approach for parallel waveform generation is to use generative adversarial networks (GANs)~\cite{goodfellow-2014-generative}.
A GAN is a powerful generative model that has been successfully used in various research fields such as image generation~\cite{reed-2016-generative}, speech synthesis~\cite{saito-2018-statistical}, and singing voice synthesis~\cite{hono-2019-singing}.
GAN-based models have also been proposed for waveform generation~\cite{yamamoto-2020-parallel,kumar-2019-melgan}.
Since these training frameworks enable models to effectively capture the time-frequency distribution of the speech waveform and improve training stability, these GAN-based models are much easier to train than the conventional non-AR methods described above.

A neural vocoder can generate high-fidelity waveforms since it can restore missing information from the acoustic feature in a data-driven fashion and is less limited by the knowledge and assumptions of the conventional vocoder~\cite{kawahara-1999-restructuring,morise-2016-world}.
However, this also results in a lack of acoustic controllability and robustness.
In fact, it is difficult for a neural vocoder to generate a speech waveform with accurate pitches outside the range of the training data.
Some methods with explicit periodic signals~\cite{wang-2019-neural,oura-2019-deep} and methods with a pitch-dependent convolution mechanism~\cite{wu-2020-quasi-pwg} address this problem.

It is known that both periodic and aperiodic components are mixed in speech waveforms.
Although neural vocoders often model speech waveforms as single signals without considering these mixed components, it is important to take them into account to model speech waveforms more effectively.
In particular, when the neural vocoder is used in a singing voice synthesis system~\cite{nishimura-2016-singing,hono-2018-recent}, the accuracy of pitch and breath sound reproduction has a significant effect on quality and naturalness.
Several methods for decomposing the periodic and aperiodic components contained in the speech waveform have been proposed~\cite{serra-1990-spectral,zubrycki-2007-accurate}.
However, it is still difficult to decompose them, and it is not optimal to use decomposed waveforms, including decomposition errors, as the training data for the neural vocoders.

In this paper, we consider speech waveform modeling in terms of the model structures and propose PeriodNet, a non-autoregressive neural vocoder for better speech waveform modeling.
We introduce two versions with different model structures, a parallel model and a series model, assuming that the periodic and aperiodic waveforms can be generated from the explicit periodic and aperiodic signals, such as a sine wave and a noise sequence, respectively.
Our proposed methods also can generate a waveform that includes a pitch outside the range of the training data.
Moreover, our models have the robustness of the input pitch since these generate the periodic and aperiodic waveforms with two separate neural networks.

\section{Waveform modeling}
\vspace{-1mm}

\subsection{Autoregressive neural vocoder}
\vspace{-1mm}
In neural vocoders with an AR structure~\cite{oord-2016-wavenet,kalchbrenner-2018-efficient,valin-2019-lpcnet}, a speech waveform at each timestep is modeled as a probability distribution conditioned on past speech samples and auxiliary features such as Mel-spectrograms and acoustic features.
An overview of the AR neural vocoder is shown in Fig.~\ref{fig:ar}.
In this paper, we use WaveNet~\cite{oord-2016-wavenet} as a neural network to generate waveforms (referred to as a generator in this paper).
WaveNet has a stack of dilated causal convolution with a gated activation function, and it is capable of modeling speech waveforms with complex periodicity.
However, there is a problem that it cannot make a parallel inference and takes time to generate waveform because of the AR structure.

\subsection{Non-autoregressive neural vocoder}
\vspace{-1mm}
In non-AR neural vocoders~\cite{oord-2018-parallel,ping-2018-clarinet,prenger-2019-waveglow,kim-2018-flowavenet,yamamoto-2020-parallel,kumar-2019-melgan}, the neural network represents the mapping function from a pre-generated input signal, such as Gaussian noise, to the speech waveform.
Hence, all waveform samples can be generated in parallel without incurring the expense of having to make predictions autoregressively.
However, it is difficult to predict a speech waveform with autocorrelation from a noise sequence without autocorrelation properly.
Prior studies~\cite{wang-2019-neural,oura-2019-deep} have proposed methods that use explicit periodic signals such as sine waves.
These methods provide high pitch accuracy and can synthesize waveforms with a pitch not included in the training data.
In this paper, following these attempts, we use the sine wave, noise, and voiced/unvoiced (V/UV) sequence as input signals, as shown in Fig~\ref{fig:baseline}.
Note that this V/UV sequence is smoothed in advance.
Various architectures can be used for the generator in Fig~\ref{fig:baseline}; we use a Parallel WaveGAN~\cite{yamamoto-2020-parallel}-based architecture.
Details of model architectures will be described in Sec.~\ref{sec:details}.

\section{Proposed model structures separating periodic and aperiodic components}
\label{sec:methods}
\vspace{-1mm}

\subsection{Model structures}
\label{sec:structures}
\vspace{-1mm}

A speech waveform contains periodic and aperiodic waveforms.
In the structure shown in Fig.~\ref{fig:baseline}, the generation process of the periodic and aperiodic waveforms is represented by a single model.
However, this structure is not always optimal for waveform modeling, especially when the accuracy of pitch and breath sound reproduction has a significantly affects quality and naturalness, such as in singing voice synthesis.
We assume that the speech waveform is the sum of periodic and aperiodic components.
The periodic and aperiodic components are expected to be easily created from the periodic and aperiodic signal (such as the sine waves and noise sequences), respectively.
Thus, in this paper, we propose a parallel mode structure and a series model structure based on these assumptions.

The parallel model structure is shown in Fig~\ref{fig:parallel}.
This structure assumes that the periodic and aperiodic waveforms are independent of each other.
An explicit periodic signal consisting of a sine wave and V/UV sequence is used to predict the periodic waveform, and an explicit aperiodic signal consisting of noise and V/UV sequence is used to predict the aperiodic waveform.

The series model structure is shown in Fig~\ref{fig:series}.
In this structure, we assume that the aperiodic waveform depends on the periodic waveform, considering the possibility that there is an aperiodic waveform corresponding to the phase of the periodic waveform.
Specifically, we introduce a residual connection between two generators so that the latter generator can predict the aperiodic component taking into account the dependence of the periodic component.

In the parallel model and the series model, different acoustic features can be selected for the auxiliary features of the periodic and aperiodic generators, making it possible to obtain more robust neural vocoders with proper conditioning.

\subsection{Model details}
\label{sec:details}
\vspace{-1mm}

\begin{figure}[t]
  \begin{minipage}{0.49\hsize}
    \centering
    \includegraphics[width=0.95\hsize]{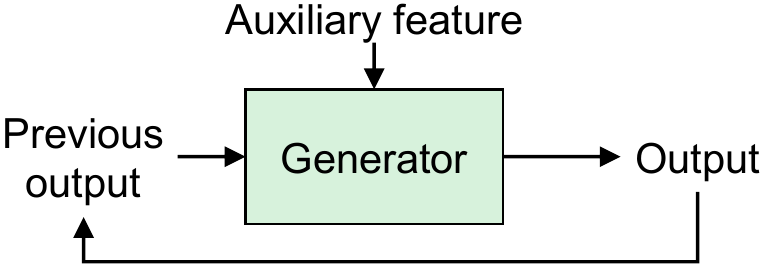}
    \subcaption{AR model}
    \label{fig:ar}
    \vspace{1mm}
  \end{minipage}
  \begin{minipage}{0.49\hsize}
    \centering
    \includegraphics[width=0.95\hsize]{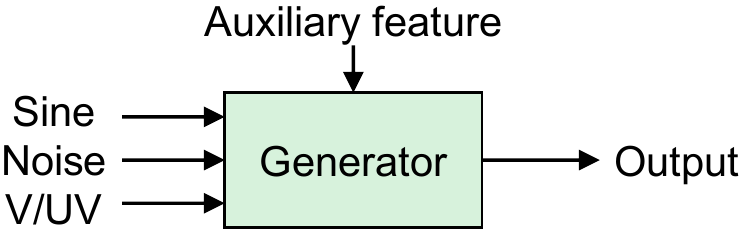}
    \subcaption{Non-AR baseline model}
    \label{fig:baseline}
  \end{minipage}
  \begin{minipage}{0.49\hsize}
    \centering
    \includegraphics[width=0.95\hsize]{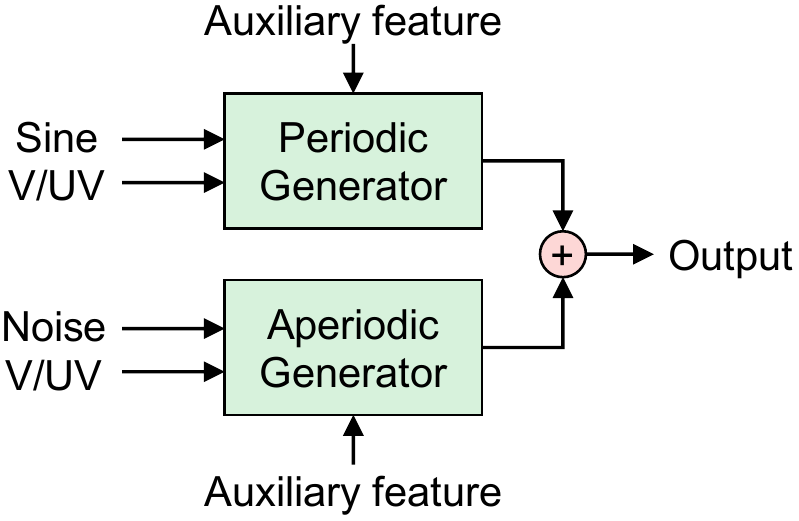}
    \subcaption{Non-AR parallel model}
    \label{fig:parallel}
  \end{minipage}
  \begin{minipage}{0.49\hsize}
    \centering
    \includegraphics[width=0.95\hsize]{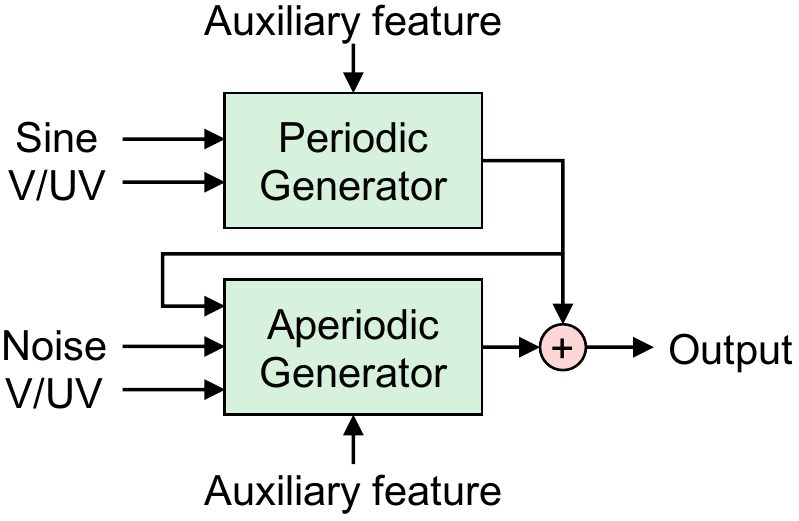}
    \subcaption{Non-AR series model}
    \label{fig:series}
  \end{minipage}
  \vspace{-2mm}
  \caption{Structures for speech waveform modeling}
  \vspace{-2mm}
  \label{fig:model}
\end{figure}

In this paper, we incorporate Parallel WaveGAN~\cite{yamamoto-2020-parallel}-based framework into our non-AR baseline and proposed models, as shown in Fig.~\ref{fig:baseline}, Fig.~\ref{fig:parallel}, and Fig.~\ref{fig:series}.
Each generator has the same architecture as the generator of \cite{yamamoto-2020-parallel}, which is a modified WaveNet-based model with non-causal convolution.
On the other hand, for the discriminators, we utilize a multi-scale architecture with three discriminators that have identical network structures but operate on different audio scales, following \cite{kumar-2019-melgan}.
Each discriminator has the same architecture as the discriminator of \cite{yamamoto-2020-parallel}.
These models are trained by optimizing the combination of multi-resolution short-time Fourier transform loss and adversarial loss in the same fashion as \cite{yamamoto-2020-parallel}.

In the training vocoder with the parallel and series model structures, the final output sequence, which is the sum of two generators' output sequence, is only evaluated.
This is the same as the baseline model with the single model structure.
From the assumptions presented in Sec.~\ref{sec:structures}, by inputting the sine wave and noise sequence separately, each generator should be trained to predict periodic and aperiodic waveforms, respectively.

\section{Experiments}
\vspace{-1mm}

\subsection{Experimental conditions}
\vspace{-1mm}

\begin{figure*}[t]
  \begin{minipage}{0.33\hsize}
    \centering
    \includegraphics[width=0.95\hsize]{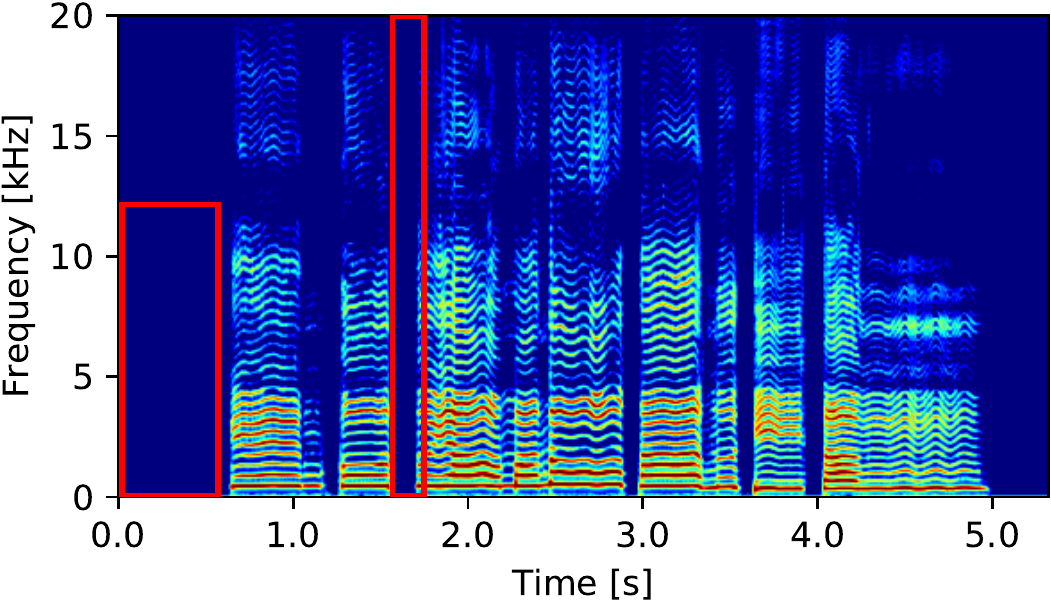}
    \vspace{-1mm}
    \subcaption{Waveform of the periodic generator's output}
    \label{fig:pm1_p}
  \end{minipage}
  \begin{minipage}{0.33\hsize}
    \centering
    \includegraphics[width=0.95\hsize]{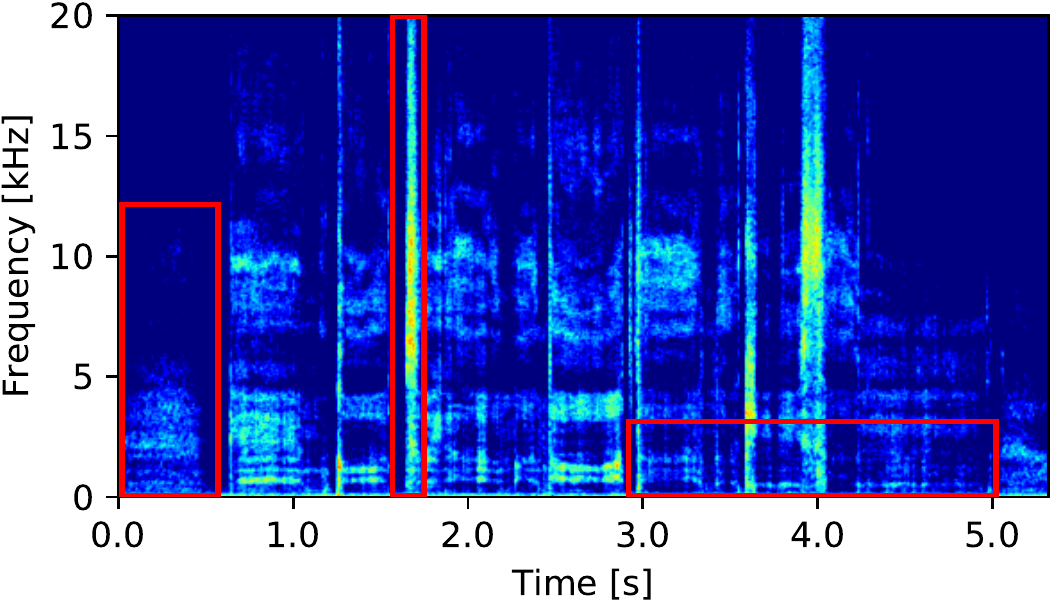}
    \vspace{-1mm}
    \subcaption{Waveform of the aperiodic generator's output}
    \label{fig:pm1_ap}
  \end{minipage}
  \begin{minipage}{0.33\hsize}
    \centering
    \includegraphics[width=0.95\hsize]{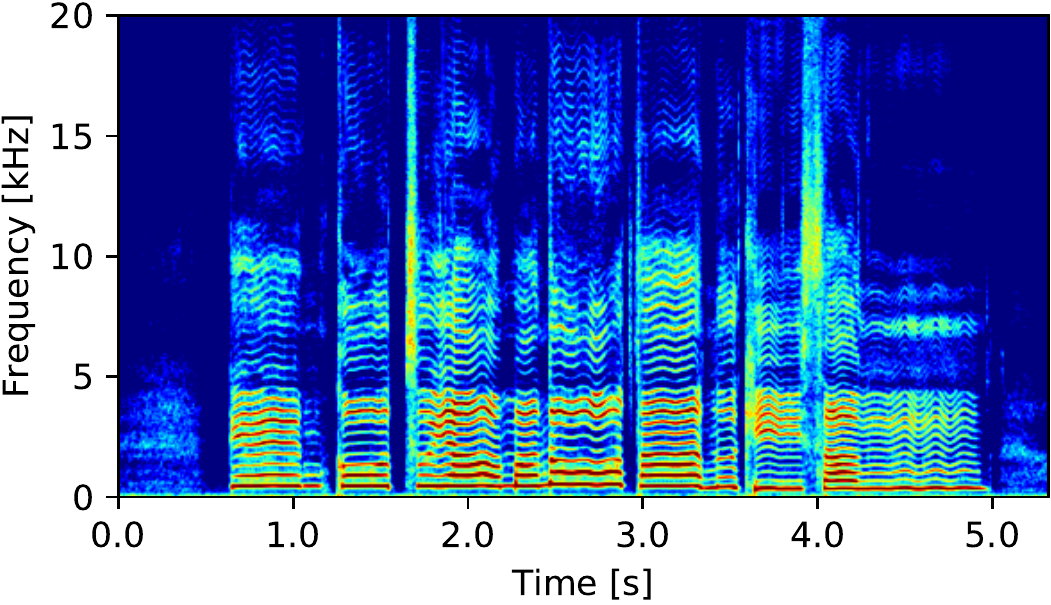}
    \vspace{-1mm}
    \subcaption{Waveform after the sum of two signals}
    \label{fig:pm1_final}
  \end{minipage}
  \vspace{-2mm}
  \caption{Spectrograms of generated waveform by non-AR parallel model}
  \label{fig:pm_spec}
\end{figure*}

\begin{figure*}[t]
  \begin{minipage}{0.33\hsize}
    \centering
    \includegraphics[width=0.95\hsize]{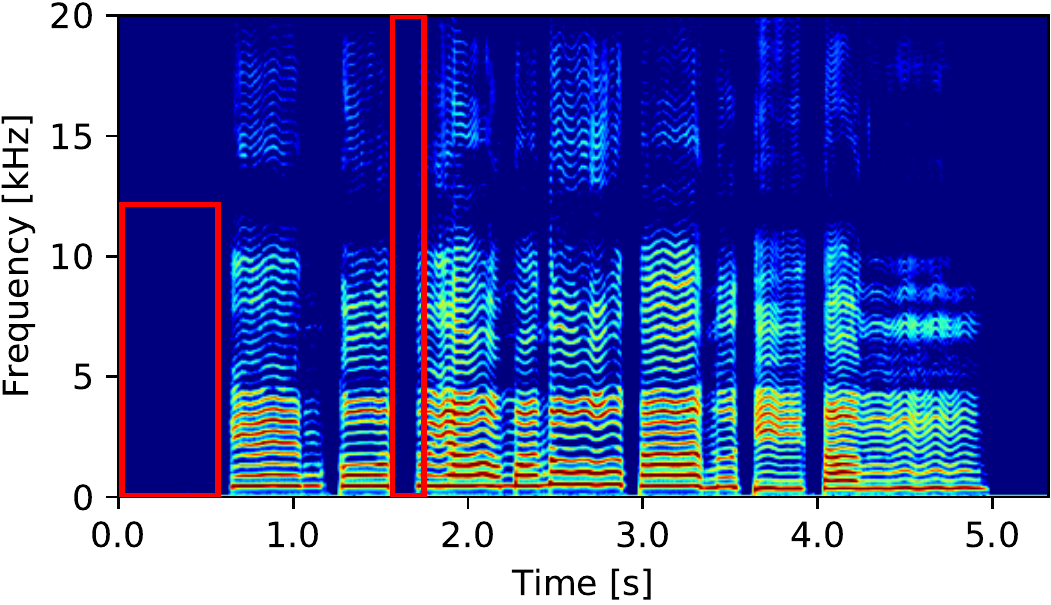}
    \vspace{-1mm}
    \subcaption{Waveform of the periodic generator's output}
    \label{fig:sm_p}
  \end{minipage}
  \begin{minipage}{0.33\hsize}
    \centering
    \includegraphics[width=0.95\hsize]{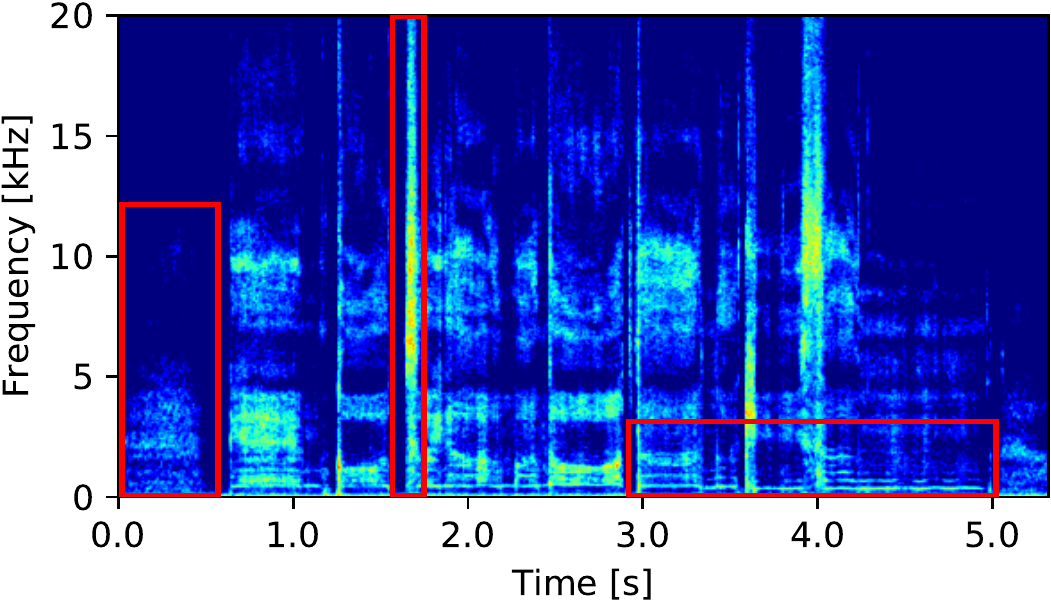}
    \vspace{-1mm}
    \subcaption{Waveform of the aperiodic generator's output}
    \label{fig:sm_ap}
  \end{minipage}
  \begin{minipage}{0.33\hsize}
    \centering
    \includegraphics[width=0.95\hsize]{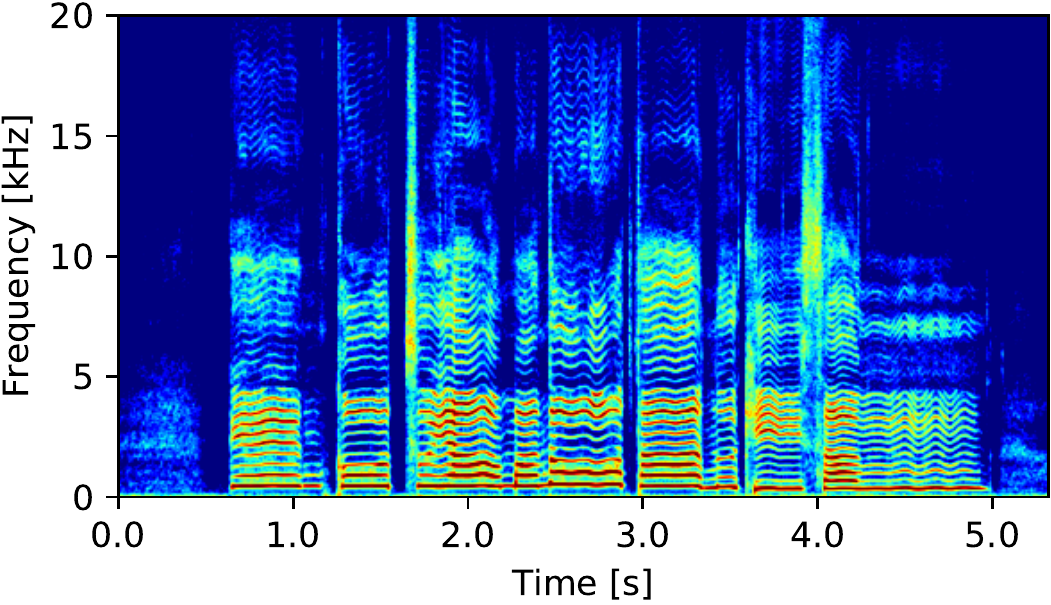}
    \vspace{-1mm}
    \subcaption{Waveform after the sum of two signals}
    \label{fig:sm_final}
  \end{minipage}
  \vspace{-2mm}
  \caption{Spectrograms of generated waveform by non-AR series model}
  \label{fig:sm_spec}
\end{figure*}

Seventy Japanese children's songs (total: 70 min) performed by one female singer were used for the experiments.
Sixty songs were used for training, and the rest were used for testing.
Singing voice signals were sampled at 48 kHz, and each sample was quantized by 16 bits.
The auxiliary features consisted of 50-dimensional WORLD mel-cepstral coefficients~\cite{morise-2016-world}, 25-dimensional mel-cepstral analysis aperiodicity measures, one-dimensional continuous log fundamental frequency ($F_0$) value, and one-dimensional voiced/unvoiced binary code.
Feature vectors were extracted with a 5-ms shift, and the features were normalized to have zero mean and unit variance before training.

In the training stage, the sine waves for the input of the non-AR neural vocoder were generated based on the glottal closure point extracted from a natural speech using REAPER~\cite{Web-REAPER}.
The purpose of this is to input a sine wave that is close in phase to the target's natural speech during training.
Meanwhile, the sine waves were generated based on the $F_0$ values in the synthesis stage.

The following seven systems were compared.
\begin{itemize}
  \item \textbf{WN}:\;The AR WaveNet~\cite{oord-2016-wavenet}.
  \item \textbf{BM1}:\;The non-AR baseline model, as shown in Fig.~\ref{fig:baseline} that used noise and a V/UV signal as the generator input and is conditioned on all auxiliary features.
  \item \textbf{BM2}:\;The non-AR baseline model, as shown in Fig.~\ref{fig:baseline} that used a sine wave and a V/UV signal as the generator input and is conditioned on all auxiliary features.
  \item \textbf{BM3}:\;The non-AR baseline model, as shown in Fig.~\ref{fig:baseline} that used noise, a sine wave, and a V/UV signal as the generator input and is conditioned on all auxiliary features.
  \item \textbf{PM1}:\;The non-AR parallel model, as shown in Fig.~\ref{fig:parallel}.
  The periodic generator takes a sine wave and a V/UV signal as input, and the aperiodic generator takes noise and a V/UV signal as input.
  Both generators are conditioned on all auxiliary features.
  \item \textbf{PM2}:\;The non-AR parallel model, as shown in Fig.~\ref{fig:parallel}.
  Unlike \textbf{PM1}, the aperiodic generator is conditioned by auxiliary features other than $F_0$.
  \item \textbf{SM}:\;The non-AR series model, as shown in Fig.~\ref{fig:series}.
  The periodic generator takes a sine wave and a V/UV signal as input, and the aperiodic generator takes noise, a V/UV signal, and the output signal of the periodic generator as input.
  Both generators are conditioned on all auxiliary features.
\end{itemize}

\textbf{WN} consisted of 30 layers of dilated residual convolution blocks with causal convolution.
The dilations of \textbf{WN} were set to $1, 2, 4, \ldots, 512$, and the 10 dilation layers were stacked three times.
The channel size for dilation, residual block, and skip-connection in \textbf{WN} was set to 256, and the filter size in \textbf{WN} was set to two.
The singing voice waveforms to train \textbf{WN} were quantized from 16 bits to 8 bits by using the $\mu$-law algorithm~\cite{recommendation-1988-pulse}.

The generators of \textbf{BM1}, \textbf{BM2}, and \textbf{BM3}, and periodic generator of \textbf{PM1}, \textbf{PM2}, and \textbf{SM} consisted of 30 layers of dilated residual convolution blocks with three dilation cycles, the same as \textbf{WN}.
The aperiodic generators of \textbf{PM1}, \textbf{PM2}, and \textbf{SM} consisted of 10 layers of dilated residual convolution blocks without dilation cycles.
The channel size for dilation, residual block, and skip-connection was set to 64, and the filter size was set to three.
The discriminators of \textbf{BM1}, \textbf{BM2}, \textbf{BM3}, \textbf{PM1}, \textbf{PM2}, and \textbf{SM} had the multi-scale architecture with three discriminators.
The discriminators took 48 kHz full-resolution waveforms, and 24 kHz and 16 kHz downsampled waveforms.
The downsampling was performed using average pooling.
Each discriminator consisted of 10 non-causal dilated convolutions with leaky ReLU activation function.
We applied weight normalization~\cite{salimans-2016-weight} to all convolutional layers.

All models were trained using the RAdam optimizer~\cite{liu-2019-radam} with 1000K iterations.
Specifically, in \textbf{BM1}, \textbf{BM2}, \textbf{BM3}, \textbf{PM1}, \textbf{PM2}, and \textbf{SM}, the discriminators were fixed for the first 100K iterations, and then both the generator and discriminator were jointly trained afterward.

\subsection{Comparison of spectrograms}
\label{sec:exp_pap}
\vspace{-1mm}

Fig.~\ref{fig:pm_spec} and Fig.~\ref{fig:sm_spec} show the spectrograms in \textbf{PM1} and \textbf{SM}, respectively.
Each figure has three spectrograms of the waveform of the periodic generator's output, the aperiodic generator's output, and the sum of two predicted signals.
Fig.~\ref{fig:pm1_p} and Fig.~\ref{fig:pm1_ap} show that the waveform of the periodic generator contains many harmonic components, and that of the aperiodic generator contains the other frequency components.
As seen in the highlighted boxes on the left and in the center, which represent parts of the breath and unvoiced plosives ``/t/'', respectively, it can be seen that the spectra of these unvoiced sounds only appear in the output of the aperiodic generator.
These tendencies can also be seen in Fig.~\ref{fig:sm_p} and Fig.~\ref{fig:sm_ap}.
These results indicate that two generators in the parallel model and the series model work on modeling the transformation from the sine waves and the noise sequence to the periodic and the aperiodic waveforms.
Comparing the highlighted box in the lower right of Fig.~\ref{fig:pm1_ap} and Fig.~\ref{fig:sm_ap}, the output waveform of the aperiodic generator in \textbf{SM} contains more harmonic components than in \textbf{PM1}.
This suggests that the periodic waveform as the input of the aperiodic generator in \textbf{SM} may have leaked into to the output of the aperiodic generator because the output waveform of the periodic generator fed into the aperiodic generator.
It should be noted that some harmonic components are also included in Fig.~\ref{fig:pm1_ap} since the periodic and aperiodic waveforms were not explicitly decomposed in the training stage.

\subsection{Subjective evaluations}
\vspace{-1mm}
\begin{figure}[t]
  \centering
  \includegraphics[height=3.8cm]{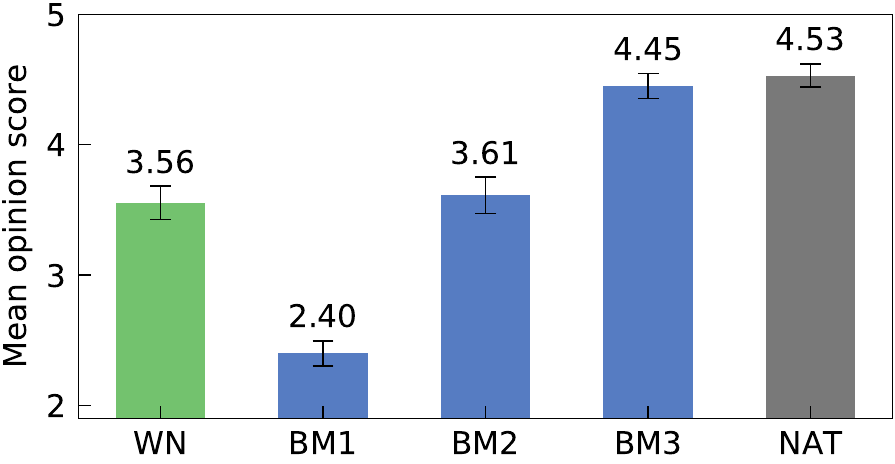}
  \vspace{-2mm}
  \caption{Subjective evaluation results of experiment 1}
  \label{fig:mos1}
\end{figure}
\begin{figure}[t]
  \centering
  \includegraphics[height=3.8cm]{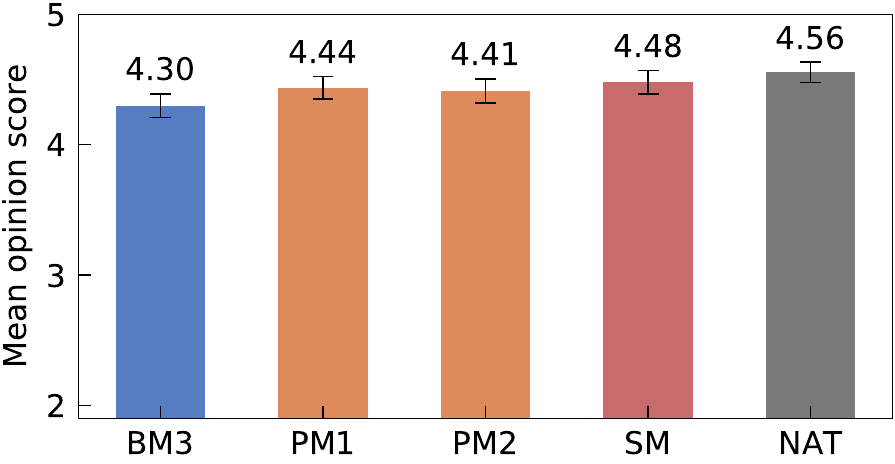}
  \vspace{-2mm}
  \caption{Subjective evaluation results of experiment 2}
  \label{fig:mos2}
\end{figure}
\begin{figure}[t]
  \centering
  \includegraphics[height=3.8cm]{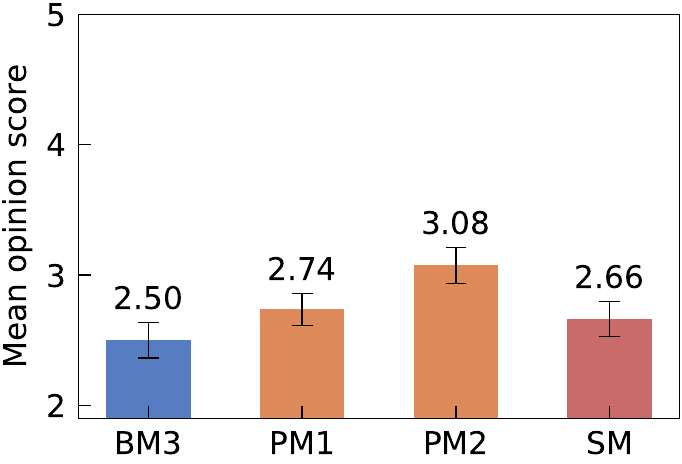}
  \vspace{-2mm}
  \caption{Subjective evaluation results of experiment 3}
  \label{fig:mos3}
\end{figure}

\subsubsection{Comparison of AR/non-AR neural vocoders and the input signals}
\vspace{-1mm}

We conducted a listening test using \textbf{WN}, \textbf{BM1}, \textbf{BM2}, \textbf{BM3}, and \textbf{NAT} to compare neural vocoders the with and without the AR structure and the input signals for the non-AR neural vocoder.
Note that \textbf{NAT} indicates a recorded natural waveform.
The naturalness of the synthesized singing voice was assumed using the mean opinion score (MOS) test method.
The participants were sixteen native Japanese speakers, and each participant evaluated ten phrases randomly selected from the test data.
After listening to each test sample in the MOS test, the participants were asked to score the naturalness of the sample out of five (1 = Bad; 2 = Poor; 3 = Fair; 4 = Good; and 5 = Excellent).

The results of the subjective evaluation are shown in Fig.~\ref{fig:mos1}.
\textbf{BM1} yielded a lower MOS value than \textbf{WN}, indicating that it is difficult to generate high-quality singing voices from noise.
On the other hand, \textbf{BM2} showed the same score as \textbf{WN}.
By inputting a periodic signal, the neural vocoder can appropriately synthesize waveforms with periodicity the lack of the AR structure.
However, the waveform of \textbf{WN} contains quantization noise, so the quality of \textbf{BM2} was insufficient.
\textbf{BM3}, which inputs both explicit periodic and aperiodic signals, has reached the MOS value close to \textbf{NAT}.
This indicates the effectiveness of using both explicit periodic and aperiodic signals as inputs for non-AR neural vocoders.

\subsubsection{Comparison of model structures of non-AR neural vocoders}
\vspace{-1mm}

To compare the model structures of non-AR neural vocoders, we conducted two subjective evaluation experiments using \textbf{BM3}, \textbf{PM1}, \textbf{PM2}, and \textbf{SM}.
In these experiments, the samples were generated by four vocoders conditioned on two different F0 scales: original and double scale.
In the experiment with the original $F_0$ scale, we also used the natural waveform \textbf{NAT} for comparison.

The results are presented in Fig.~\ref{fig:mos2} and Fig.~\ref{fig:mos3}.
These figures show that \textbf{PM1}, \textbf{PM2}, and \textbf{SM} attained higher naturalness than \textbf{BM3}.
This indicates that it is effective for the non-AR neural vocoders using the explicit periodic signal to introduce a parallel or series structure.
Although the difference between \textbf{PM1}, \textbf{PM2}, and \textbf{SM} was negligible when conditioning on the original $F_0$ as shown in Fig.~\ref{fig:mos2}, \textbf{PM2} was the best performance when conditioning on the doubled $F_0$ as shown in Fig.~\ref{fig:mos3}.
The waveform samples generated by \textbf{BM3}, \textbf{PM1}, and \textbf{SM} tended to contain more aperiodic waveforms than those generated by \textbf{PM2}.
In \textbf{BM3}, the period and aperiodic components were not modeled separately, and speech waveforms were generated from a single generator conditioned by auxiliary features including $F_0$.
In \textbf{PM1} and \textbf{SM}, although the networks for modeling these components were separate, both aperiodic generators were conditioned on auxiliary features including $F_0$.
In particular, the aperiodic generator in \textbf{SM} also depended on periodic waveforms predicted by the periodic generator.
Therefore, it was assumed that \textbf{BM3}, \textbf{PM1}, and \textbf{SM} could not generate aperiodic waveforms when these vocoders took out-of-range $F_0$ as the acoustic features in the synthesis stage.
\textbf{PM2} is more robust for an unseen $F_0$ outside the $F_0$ range of the training data because the aperiodic generator in \textbf{PM2} does not depend on the periodic signal or $F_0$.

\section{Conclusions}
\vspace{-1mm}
We introduced PeriodNet, a non-AR neural vocoder with new model structures, to appropriately modeling the periodic and aperiodic components in the speech waveform.
Each generator in the parallel or series model structure can model the periodic and aperiodic waveforms without the use of decomposition techniques.
The experimental results showed that the proposed methods were able to generate high-fidelity speech waveforms and improve the ability to generate waveforms with a pitch outside the range of the training data.
Future work includes investigating the effect of proposed methods on different datasets, such as a multi-speaker and multi-singer dataset.

\section{Acknowledgements}
\vspace{-1mm}
This work was supported by JSPS KAKENHI Grant Number JP19H04136 and JP18K11163.

\bibliographystyle{IEEEtran}
\bibliography{paper}

\begin{thebibliography}{10}
\providecommand{\url}[1]{#1}
\csname url@samestyle\endcsname
\providecommand{\newblock}{\relax}
\providecommand{\bibinfo}[2]{#2}
\providecommand{\BIBentrySTDinterwordspacing}{\spaceskip=0pt\relax}
\providecommand{\BIBentryALTinterwordstretchfactor}{4}
\providecommand{\BIBentryALTinterwordspacing}{\spaceskip=\fontdimen2\font plus
\BIBentryALTinterwordstretchfactor\fontdimen3\font minus
  \fontdimen4\font\relax}
\providecommand{\BIBforeignlanguage}[2]{{%
\expandafter\ifx\csname l@#1\endcsname\relax
\typeout{** WARNING: IEEEtran.bst: No hyphenation pattern has been}%
\typeout{** loaded for the language `#1'. Using the pattern for}%
\typeout{** the default language instead.}%
\else
\language=\csname l@#1\endcsname
\fi
#2}}
\providecommand{\BIBdecl}{\relax}
\BIBdecl

\bibitem{oord-2016-wavenet}
A.~van~den Oord, S.~Dieleman, H.~Zen, K.~Simonyan, O.~Vinyals, A.~Graves,
  N.~Kalchbrenner, A.~W. Senior, and K.~Kavukcuoglu, ``Wave{N}et: {A}
  generative model for raw audio,'' \emph{arXiv preprint arXiv:1609.03499},
  2016.

\bibitem{tamamori-2017-speaker}
A.~Tamamori, T.~Hayashi, K.~Kobayashi, K.~Takeda, and T.~Toda,
  ``Speaker-dependent {W}ave{N}et vocoder.'' in \emph{Proccdings of
  Interspeech}, 2017, pp. 1118--1122.

\bibitem{shen-2018-natural}
J.~Shen, R.~Pang, R.~J. Weiss, M.~Schuster, N.~Jaitly, Z.~Yang, Z.~Chen,
  Y.~Zhang, Y.~Wang, R.~Skerrv-Ryan \emph{et~al.}, ``Natural {TTS} synthesis by
  conditioning {W}ave{N}et on mel spectrogram predictions,'' in
  \emph{Proceedings of ICASSP}, 2018, pp. 4779--4783.

\bibitem{ping-2017-deep}
W.~Ping, K.~Peng, A.~Gibiansky, S.~O. Arik, A.~Kannan, S.~Narang, J.~Raiman,
  and J.~Miller, ``Deep voice 3: Scaling text-to-speech with convolutional
  sequence learning,'' \emph{arXiv preprint arXiv:1710.07654}, 2017.

\bibitem{kalchbrenner-2018-efficient}
N.~Kalchbrenner, E.~Elsen, K.~Simonyan, S.~Noury, N.~Casagrande, E.~Lockhart,
  F.~Stimberg, A.~v.~d. Oord, S.~Dieleman, and K.~Kavukcuoglu, ``Efficient
  neural audio synthesis,'' \emph{arXiv preprint arXiv:1802.08435}, 2018.

\bibitem{valin-2019-lpcnet}
J.-M. Valin and J.~Skoglund, ``{LPCNet}: Improving neural speech synthesis
  through linear prediction,'' in \emph{Proceedings of ICASSP}, 2019, pp.
  5891--5895.

\bibitem{oord-2018-parallel}
A.~van~den Oord, Y.~Li, I.~Babuschkin, K.~Simonyan, O.~Vinyals, K.~Kavukcuoglu,
  G.~Driessche, E.~Lockhart, L.~Cobo, F.~Stimberg \emph{et~al.}, ``Parallel
  {W}ave{N}et: Fast high-fidelity speech synthesis,'' in \emph{Proceedings of
  ICML}, 2018, pp. 3918--3926.

\bibitem{ping-2018-clarinet}
W.~Ping, K.~Peng, and J.~Chen, ``{C}lari{N}et: Parallel wave generation in
  end-to-end text-to-speech,'' \emph{arXiv preprint arXiv:1807.07281}, 2018.

\bibitem{kingma-2016-improved}
D.~P. Kingma, T.~Salimans, R.~Jozefowicz, X.~Chen, I.~Sutskever, and
  M.~Welling, ``Improved variational inference with inverse autoregressive
  flow,'' in \emph{Advances in Neural Information Processing Systems}, 2016,
  pp. 4743--4751.

\bibitem{prenger-2019-waveglow}
R.~Prenger, R.~Valle, and B.~Catanzaro, ``Wave{G}low: A flow-based generative
  network for speech synthesis,'' in \emph{Proceedings of ICASSP}, 2019, pp.
  3617--3621.

\bibitem{kim-2018-flowavenet}
S.~Kim, S.-g. Lee, J.~Song, J.~Kim, and S.~Yoon, ``Flo{W}ave{N}et: A generative
  flow for raw audio,'' \emph{arXiv preprint arXiv:1811.02155}, 2018.

\bibitem{goodfellow-2014-generative}
I.~Goodfellow, J.~Pouget-Abadie, M.~Mirza, B.~Xu, D.~Warde-Farley, S.~Ozair,
  A.~Courville, and Y.~Bengio, ``Generative adversarial nets,'' in
  \emph{Advances in Neural Information Processing Systems}, 2014, pp.
  2672--2680.

\bibitem{reed-2016-generative}
S.~Reed, Z.~Akata, X.~Yan, L.~Logeswaran, B.~Schiele, and H.~Lee, ``Generative
  adversarial text to image synthesis,'' \emph{Proceedings of ICML}, vol.~48,
  pp. 1060--1069, 20--22 Jun 2016.

\bibitem{saito-2018-statistical}
Y.~Saito, S.~Takamichi, and H.~Saruwatari, ``Statistical parametric speech
  synthesis incorporating generative adversarial networks,'' \emph{IEEE/ACM
  Transactions on Audio, Speech, and Language Processing}, vol.~26, no.~1, pp.
  84--96, 2018.

\bibitem{hono-2019-singing}
Y.~Hono, K.~Hashimoto, K.~Oura, Y.~Nankaku, and K.~Tokuda, ``Singing voice
  synthesis based on generative adversarial networks,'' in \emph{Proceedings of
  ICASSP}, 2019, pp. 6955--6959.

\bibitem{yamamoto-2020-parallel}
R.~Yamamoto, E.~Song, and J.-M. Kim, ``Parallel {W}ave{GAN}: A fast waveform
  generation model based on generative adversarial networks with
  multi-resolution spectrogram,'' in \emph{Proceedings of ICASSP}, 2020, pp.
  6199--6203.

\bibitem{kumar-2019-melgan}
K.~Kumar, R.~Kumar, T.~de~Boissiere, L.~Gestin, W.~Z. Teoh, J.~Sotelo,
  A.~de~Br{\'e}bisson, Y.~Bengio, and A.~C. Courville, ``Mel{GAN}: Generative
  adversarial networks for conditional waveform synthesis,'' in \emph{Advances
  in Neural Information Processing Systems}, 2019, pp. 14\,910--14\,921.

\bibitem{kawahara-1999-restructuring}
H.~Kawahara, I.~Masuda-Katsuse, and A.~De~Cheveigne, ``Restructuring speech
  representations using a pitch-adaptive time--frequency smoothing and an
  instantaneous-frequency-based {F}0 extraction: Possible role of a repetitive
  structure in sounds,'' \emph{Speech communication}, vol.~27, no.~3, pp.
  187--207, 1999.

\bibitem{morise-2016-world}
M.~Morise, F.~Yokomori, and K.~Ozawa, ``{WORLD}: A vocoder-based high-quality
  speech synthesis system for real-time applications,'' \emph{IEICE
  Transactions on Information and Systems}, vol.~99, no.~7, pp. 1877--1884,
  2016.

\bibitem{wang-2019-neural}
X.~Wang, S.~Takaki, and J.~Yamagishi, ``Neural source-filter waveform models
  for statistical parametric speech synthesis,'' \emph{IEEE/ACM Transactions on
  Audio, Speech, and Language Processing}, vol.~28, pp. 402--415, 2019.

\bibitem{oura-2019-deep}
K.~Oura, K.~Nakamura, K.~Hashimoto, Y.~Nankaku, and K.~Tokuda, ``Deep neural
  network based real-time speech vocoder with periodic and aperiodic inputs,''
  in \emph{Procedings of ISCA SSW10}, 2019, pp. 13--18.

\bibitem{wu-2020-quasi-pwg}
Y.-C. Wu, T.~Hayashi, T.~Okamoto, H.~Kawai, and T.~Toda, ``Quasi-periodic
  {P}arallel {W}ave{GAN}: A non-autoregressive raw waveform generative model
  with pitch-dependent dilated convolution neural network,'' \emph{arXiv
  preprint arXiv:2007.12955}, 2020.

\bibitem{nishimura-2016-singing}
M.~Nishimura, K.~Hashimoto, K.~Oura, Y.~Nankaku, and K.~Tokuda, ``Singing voice
  synthesis based on deep neural networks,'' in \emph{Proccdings of
  Interspeech}, 2016, pp. 2478--2482.

\bibitem{hono-2018-recent}
Y.~Hono, S.~Murata, K.~Nakamura, K.~Hashimoto, K.~Oura, Y.~Nankaku, and
  K.~Tokuda, ``Recent development of the {DNN}-based singing voice synthesis
  system -- {S}insy,'' in \emph{Proceedings of APSIPA}, 2018, pp. 1003--1009.

\bibitem{serra-1990-spectral}
X.~Serra and J.~Smith, ``Spectral modeling synthesis: A sound
  analysis/synthesis system based on a deterministic plus stochastic
  decomposition,'' \emph{Computer Music Journal}, vol.~14, no.~4, pp. 12--24,
  1990.

\bibitem{zubrycki-2007-accurate}
P.~Zubrycki and A.~Petrovsky, ``Accurate speech decomposition into periodic and
  aperiodic components based on discrete harmonic transform,'' in
  \emph{Proceedings of European Signal Processing Conference}, 2007, pp.
  2336--2340.

\bibitem{Web-REAPER}
``{REAPER}: Robust epoch and pitch estimator,''
  \url{https://github.com/google/REAPER}.

\bibitem{recommendation-1988-pulse}
``Pulse code modulation ({PCM}) of voice frequencies,'' in \emph{ITU-T
  Recommendation G.711}, 1988.

\bibitem{salimans-2016-weight}
T.~Salimans and D.~P. Kingma, ``Weight normalization: A simple
  reparameterization to accelerate training of deep neural networks,'' in
  \emph{Advances in Neural Information Processing Systems}, 2016, pp. 901--909.

\bibitem{liu-2019-radam}
L.~Liu, H.~Jiang, P.~He, W.~Chen, X.~Liu, J.~Gao, and J.~Han, ``On the variance
  of the adaptive learning rate and beyond,'' in \emph{Proceedings of ICLR},
  April 2020.

\end{thebibliography}

\end{document}